%% file: main_arxiv.tex
\documentclass[12pt,a4paper]{article}
\usepackage[utf8]{inputenc}
\usepackage[T1]{fontenc}
\usepackage{lmodern}
\usepackage{amsfonts}
\usepackage{amsmath}
\usepackage[pdftex]{graphicx} 
\usepackage[pdftex,linkcolor=black,pdfborder={0 0 0}]{hyperref} 
\usepackage{calc} 
\usepackage{enumitem} 
\usepackage{amsthm}
\usepackage{bm}
\usepackage{xcolor}
\usepackage{natbib}
\usepackage{placeins}
\usepackage{comment}
\usepackage{algorithm2e}
\RestyleAlgo{ruled}
\usepackage{dsfont}
\usepackage{pdfpages}
\usepackage{comment}

\def\spacingset#1{\renewcommand{\baselinestretch}%
{#1}\small\normalsize} \spacingset{1}

\input{often_used_notation}

\begin{document} 

\begin{titlepage}

\title{Predictive variational inference for flexible\\
regression models}

\author{Lucas Kock$\mbox{}^1$, Scott A.~Sisson$\mbox{}^2$, G.~S.~Rodrigues$\mbox{}^3$, and David J. Nott$\mbox{}^{1}$} 

\date{}
\maketitle
\thispagestyle{empty}
\noindent
\vspace{-2em}

\begin{center}
{\Large Abstract}
\end{center}
\vspace{-1pt}
\noindent A conventional Bayesian approach to prediction uses the 
posterior distribution to integrate out
parameters in a density for unobserved data conditional on the observed 
data and parameters.  
When the true posterior is intractable, it is replaced by an approximation; here we focus on variational approximations. Recent work has explored methods that learn posteriors optimized for predictive accuracy under a chosen scoring rule, while regularizing toward the prior or conventional posterior. Our work builds on an existing predictive variational inference (PVI) framework that improves prediction, but also diagnoses model deficiencies through implicit model expansion.
In models where the sampling density depends
on the parameters through a linear predictor, we improve the
interpretability of existing PVI methods as a diagnostic tool.  
This is achieved by adopting PVI posteriors of Gaussian mixture form (GM-PVI) and establishing connections with plug-in prediction for mixture-of-experts models. We make three main contributions. 
First, we show that GM-PVI prediction is equivalent to plug-in prediction for certain mixture-of-experts models with covariate-independent weights in generalized linear models and hierarchical extensions of them.  Second, we extend 
standard PVI by allowing GM-PVI posteriors to vary with the prediction covariate and
in this case an equivalence to plug-in prediction for mixtures of experts with covariate-dependent weights is established.   
Third, we demonstrate the diagnostic value of this approach across several examples, including generalized linear models, linear mixed models, and latent Gaussian process models, demonstrating how the parameters of the original model must vary across the covariate space to achieve improvements in prediction.

\vspace{10pt}
\noindent
{\bf Keywords}: generalized Bayes, generalised linear models, mixture models, posterior predictive, predictively oriented posteriors, variational Bayes

\vspace*{\fill}

\noindent {\small\textbf{Acknowledgments:} SAS is supported by the Australian Research Council through the Discovery Project program (DP260101134). David Nott's research was supported by the Ministry of Education, Singapore, under the Academic Research Fund Tier 2 (MOE-T2EP20123-0009), and he is affiliated with the Institute of Operations Research and Analytics at the National University of Singapore.}

\vspace{10pt}

\noindent{\small
$^1$ Department of Statistics and Data Science, National University of Singapore, Singapore; \mbox{lucas.kock@nus.edu.sg}, standj@nus.edu.sg\\
$^2$ School of Mathematics and Statistics, University of New South Wales, Australia; scott.sisson@unsw.edu.au\\
$^3$ Department of Statistics, University of Brasília, Brazil; guilhermerodrigues@unb.br\\
}

\end{titlepage}

\spacingset{1.5} 

\input{text.tex}
\FloatBarrier
\setlength{\bibsep}{0pt plus 0.3ex}
\bibliography{bib}
\FloatBarrier
\appendix
\section*{Supplementary Information}
\input{appendix.tex}
\end{document}

%% file: often_used_notation.tex
\newcommand{\ND}{\mathcal{N}}
\newcommand{\PoisD}{Pois}
\newcommand{\BernD}{\text{Bern}}
\DeclareMathOperator{\trace}{\text{tr}}

%% file: text.tex
\section{Introduction}\label{sec:Intro}

It is well-known that Bayesian inference can perform poorly in situations
where the model is misspecified.  Bayesian inference about the parameters
can become meaningless under misspecification \citep{walker13}, and 
Bayesian prediction can only be defended as optimal when the model is
correct \citep{aitchison75}.  
A growing literature has proposed modifications to conventional 
Bayesian predictive inference with improved performance 
under misspecification. Some of these modifications simultaneously pursue
another goal: providing insight into how the model can be improved. 
The idea is that an expanded model could recover the predictive gains 
offered by the modified predictive approach for the original model, 
but in a conventional Bayesian way. 
This paper focuses on so-called predictive
variational inference (PVI) methods \citep[e.g.~][]{lai+y25}.  
In PVI methods, 
a PVI posterior is learnt on the parameters by optimizing a criterion 
that leads to good prediction when parameters are integrated out using the PVI posterior in forming predictive distributions.  
We focus here on obtaining interpretable PVI posteriors
that can guide model improvement.    

To attain the goal of good prediction under misspecification 
which is informative for model expansion, 
we consider regression models where the
sampling density for a new observation depends on the unknown parameters through a 
scalar linear predictor.  Common models satisfying this requirement include
generalized linear models, their hierarchical extensions, and latent
Gaussian process models.  
We consider Gaussian mixture PVI approximations, 
and using the assumed structure of the model we can compute predictive densities
using only one-dimensional quadrature.  These predictive densities are themselves
mixtures, which follows from the Gaussian mixture PVI (GM-PVI) 
posterior form.  We make three main contributions.
First, we establish a connection between GM-PVI posterior prediction and 
mixture-of-experts model expansions with covariate-independent weights. 
Second, we establish an analogous connection for mixture-of-experts expansions 
with covariate-dependent weights, enabled by a novel extension in which the 
GM-PVI posterior is allowed to vary with the 
prediction covariate.  This yields what 
we call varying GM-PVI (VGM-PVI) posteriors. Third, through a range of examples, 
we show that the mixture-of-experts connection provides insight into 
model deficiencies, demonstrating how the parameters must change across the 
covariate space to improve prediction.
The most important contribution is the second one: allowing the PVI posterior 
to depend on covariates in regression settings. Our examples demonstrate 
that VGM-PVI substantially enhances the utility of PVI as a tool for model criticism.

Our work is most closely related to that of 
\cite{lai+y25}, although other related approaches
are discussed below.  In their PVI approach, \cite{lai+y25} 
consider a parametrized form for a posterior approximation, but instead of
approximating the conventional posterior distribution,   
which may not be desirable under model
misspecification, they optimize predictive performance of the PVI posterior
according to a scoring rule.  In the optimization, they 
regularize the PVI posterior either towards
the prior or conventional posterior.  Their work is motivational for ours in
their focus on diagnosing misspecification, and their interpretation of PVI
as an implicit model expansion which can be insightful for model improvement. 
In particular, they consider the change in variability between the PVI posterior
and conventional posterior as suggestive of which parameters in the current model
need to vary across observations to achieve a better model.  
\cite{lai+y25} describe methods for implementing PVI for several different
scoring rules, and also in the context of likelihood-free or simulation-based
inference \citep{sisson+fb18intro,cranmer+bl20}.    
The method of \cite{lai+y25} can be thought of as a continuous version
of Bayesian model stacking \citep{yao+vsg18}, where a density on 
parameters is optimized
for prediction rather than
a set of weights for a discrete set of models.  In this analogy, our
novel extension of PVI posteriors allowing them to vary with prediction covariates
is analogous to hierarchical model stacking \citep{yao+pvg22}.  

Recent work of \cite{MclCheFraKno2025} provides the most comprehensive theoretical
analysis of predictive variational and related predictively oriented 
Bayesian methods to date.  
They define a P{\footnotesize R}O posterior as a solution
to an optimization problem where the objective incorporates a scoring rule
term and a term regularizing towards the prior. 
\cite{MclCheFraKno2025} prove some results concerning the behaviour of the 
P{\footnotesize R}O posterior relative to a Gibbs posterior based on the same
scoring rule.  One of these shows that
under correct model specification the 
P{\footnotesize R}O posterior achieves the same predictive performance 
as the corresponding Gibbs posterior, while in 
non-trivial misspecified settings it is strictly 
superior in terms of the divergence induced by the scoring rule.   
They devise algorithms for computing P{\footnotesize R}O posteriors
using a mean-field 
Langevin dynamics approach; \cite{chazal+ksko25} discuss 
alternative computational methods 
that may require less tuning. 
\citet[Section~4.2.2]{MclCheFraKno2025} also discuss the 
relationship of the P{\footnotesize R}O posterior to mixture 
models for the data, which is relevant to our current work.  
Our approach differs, however, in adopting a finite mixture form for the PVI posterior itself.
  
Also relevant to our work is research on generalized Bayes and generalized
variational inference \citep{knoblauch+jd22} and PAC-Bayesian bounds
\citep{alquier24}.  
\cite{masegosa20} emphasizes the difference between
inferential and predictive risk, and bound the latter using
second-order PAC-Bayesian bounds.  Their sharper
bounds lead to improved predictive performance
according to the log score, and contain a diversity term that 
provides a novel interpretation of why
diverse ensembles are beneficial in ensemble learning.  
Related work of \cite{morningstar+ad22} develops PAC$^m$-Bayes, giving
a multi-sample bound for predictive risk based on the log score 
which becomes tight as the sample size 
tuning parameter $m$ in the bound increases.    
We also focus on the log score in our work, but in situations where
the log score can be computed in closed form, or accurately computed 
using one-dimensional quadrature.   
\cite{shen+kpo25} consider an approach to uncertainty quantification in misspecified
deterministic models, where predictive distributions are formed by placing a mixing
distribution on the model parameters. They consider a generalized variational
inference criterion involving a maximum mean discrepancy between the variational
predictive and the empirical distribution of the data, together with a 
regularization term involving Kullback-Leibler 
divergence between the mixing distribution
and the prior. Their work is novel in considering uncertainty quantification
for deterministic models, but this will not be our focus here.  

A number of authors have considered generalized variational inference
approaches to improve prediction in the particular context of 
sparse Gaussian processes \citep[GPs;][]{titsias09} 
and related models.   \cite{jankowiak+pg20a} considers 
the form of the predictive distribution arising in sparse GPs and define
a parametric model motivated by this, optimizing parameters directly based on
a log score with a regularization term.  \cite{jankowiak+pg20a} considers a related
approach, where a so-called deep sigma point process (DSPP) model is defined, motivated
by the form of the predictive distribution for a deep Gaussian process.  The DSPP
model replaces continuous mixing in the deep GP approach 
with a finite mixture based on quadrature
approximations.  The quadrature points and weights are free parameters which
are optimized for predictive performance along with other parameters by regularized
maximum likelihood.  The predictive distribution has the form of a Gaussian
mixture, which we also consider here, but we focus on different types of models 
and also consider variational posteriors which vary according to the covariate
at which prediction is required.  Several authors 
consider direct loss minimization for sparse Gaussian processes, 
where the direct loss is the log score.  
\cite{sheth+k20} derive risk bounds which apply for a smoothed log loss
in sparse GP regression.  \cite{wei+sk21} focuses on computational issues, 
addressing the issue of gradient estimation for direct loss minimization. 

There are many novel recent developments in Bayesian prediction broadly, and 
we refer the reader to a recent special issue 
in the journal {\it Statistical Science}
for an excellent overview \citep{clarke+r25}.  Since we focus on 
PVI methods which are interpretable for model expansion, our work is
also related to Bayesian predictive model checking (see, for example, 
\citealt{gelman+ms96} and \citealt[Chapter 5]{evans15}).   
However, the PVI methods here are
different to conventional model checking approaches
in focusing on modifying prediction in an existing model
to obtain better predictions in addition to understanding
misspecification.  

In the next section, we provide background on variational inference and 
Bayesian predictive inference. Section 3 develops PVI with a Gaussian 
mixture variational family, establishing connections with plug-in prediction 
for mixture-of-experts models, and introduces VGM-PVI posteriors that vary 
with the prediction covariates, linking them to mixture-of-experts expansions 
with covariate-dependent weights. Sections 4 through 6 present examples. 
Section 4 considers linear, logistic, and Poisson regression,  
Section 5 considers hierarchical models with random effects
and Section 6 considers latent Gaussian process models. 
Section 7 gives concluding discussion. 
Code is publicly available at \href{https://github.com/kocklucx/GMPVI}{github.com/kocklucx/GMPVI}.

\section{Background}\label{sec:setup}

Suppose we have a statistical model with parameters $\theta\in \Theta$ for data $y=(y_1,\dots, y_n)^\top\in \mathcal{Y}^n$.  We adopt a Bayesian approach to learning about $\theta$, using a prior density
$p(\theta)$.  Regression problems are considered where $y$ is a vector of responses, and
for each $y_i$ there is an associated covariate $x_i=(x_{i1},\dots, x_{ip})^\top\in \mathcal{X}$.  We write
$X=\{x_1,\dots, x_n\}$.  The covariates $X$ are not modelled, but only $y|X$.  
The sampling density of $y|X,\theta$ is $p(y|X,\theta)$, but henceforth we generally write 
this as $p(y|\theta)$ for simplicity.   
The posterior density is $p(\theta|y) \propto p(\theta)p(y|\theta)$.  
Similar to our notation for the sampling density, we suppress dependence on covariates in our notation for
posterior densities and predictive densities, except where
showing the dependence explicitly is necessary for clarity. 

In Bayesian inference we need to summarize the posterior distribution, 
and variational inference \citep[VI;~][]{BleKucMca2017} is one method to do this.  It optimizes
a posterior approximation in a convenient class
of densities which we denote as $\mathcal{Q}=\{q_\lambda(\theta); \lambda\in\Lambda\}$, where
$\lambda$ is a vector of variational parameters.   A commonly used measure of the quality
of a variational approximation is the Kullback-Leibler (KL) divergence, and in this case
the approximate posterior density is $q_{\lambda^*}(\theta)$, where 
\begin{align}
 \lambda^*= & \arg \min_{\lambda\in \Lambda} \text{KL}(q_{\lambda}(\theta)||p(\theta|y)),  \label{minimize}
\end{align} 
and
\begin{align}
  \text{KL}(q_\lambda(\theta)||p(\theta|y)) & = 
  \int q_\lambda(\theta)\log \frac{q_\lambda(\theta)}{p(\theta|y)}\,d\theta.
\end{align}
We have assumed a unique minimizer $\lambda^*$.  The optimization \eqref{minimize}
is equivalent to maximization of the so-called evidence lower bound \citep[ELBO;][]{OrmWan2010}, 
\begin{align}
  \lambda^* = & \arg \max_{\lambda\in\Lambda} \text{ELBO}(\lambda), \label{maximize}
\end{align}
where $\text{ELBO}(\lambda):=\mathbb{E}_q[\log p(\theta)p(y|\theta)-\log q_{\lambda}(\theta)]$ and
$\mathbb{E}_q[\cdot]$ denotes expectation with respect to $q_\lambda(\theta)$.  The ELBO gets its name
as a lower bound on the evidence $\log p(y)$, where
$$p(y)=\int p(\theta)p(y|\theta)\,d\theta.$$
The ELBO is convenient to work with in optimization because its expression 
does not involve any unknown posterior normalizing constant.  

Let
$\widetilde{y}$ denote an unobserved response with observed covariate
$\widetilde{x}\in \mathcal{X}$, and suppose the model for $\widetilde{y}|\widetilde{x}$ is 
$p(\widetilde{y}|\widetilde{x},\theta)$.  In traditional Bayesian inference, 
the predictive density for $\widetilde{y}|\widetilde{x},y$ is
\begin{align}
  p(\widetilde{y}|\widetilde{x},y) & = \int p(\widetilde{y}|\widetilde{x},\theta)p(\theta|y)\,d\theta, \label{conventional}
\end{align}
where it is assumed here that $\widetilde{y}$ and $y$ are conditionally independent given $\theta$
and the covariates.
The PVI approach of \citet{lai+y25} replaces the posterior distribution in the integrand in 
\eqref{conventional} with a PVI posterior
$q_\lambda(\theta)\in \mathcal{Q}$, 
to construct a predictive density as
\begin{align}
   q_\lambda(\widetilde{y}|\widetilde{x}):= \int p(\widetilde{y}|\widetilde{x},\theta)q_\lambda(\theta)\,d\theta.   \label{PVI}
\end{align} 
PVI chooses $\lambda$ so that the predictions obtained by \eqref{PVI} are optimal
according to some chosen scoring rule.  
Some background on scoring rules is given now;  this is necessary so that the
objective function used to choose $\lambda$ in PVI can be explained.

Let $\mathcal{P}$ be a space of distributions for data $z\in \mathcal{Y}$ to be observed.  
A scoring rule is a function $\mathcal{S}:\mathcal{P}\times \mathcal{Y}\rightarrow \overline{\mathbb{R}}$, 
where $\overline{\mathbb{R}}=\mathbb{R}\cup \{-\infty,\infty\}$.  For data $z$ and $P\in \mathcal{P}$, we think of 
$S(P,z)$ as a measure of consistency of the data $z$ with the distribution $P$.  We will 
define scoring rules so that a larger value means greater consistency (so-called positively-oriented
scoring rules).
If $P,Q\in \mathcal{P}$, we use a common abuse of notation and write
$S(P,Q):=\mathbb{E}_{Z\sim Q}(S(P,Z))$.  A scoring rule is proper for $\mathcal{P}$ if
$S(P,P)\geq S(P,Q)$ for any $Q\in \mathcal{P}$.  It is strictly proper for $\mathcal{P}$ if $S(P,P')\geq S(P,Q)$ 
for all $Q\in \mathcal{P}$ implies that $P=P'$.  If $P$ and $Q$ have densities $p(\cdot)$ and $q(\cdot)$ say, we will also write
$S(p(\cdot),z):=S(P,z)$ and $S(p(\cdot),q(\cdot)):=S(P,Q)$.  For further background on scoring
rules, see \cite{gneiting+r07}.  

For a given scoring rule, we can define the objective optimized in PVI as follows.  
We change the optimization problem \eqref{minimize} to
\begin{align}
\lambda^*=&\arg \max_{\lambda} \left\{\sum_{i=1}^n S(q_\lambda(\cdot|x_i),y_i)-\beta.
\text{KL}(q_\lambda(\theta)||p(\theta|y))\right\}, \label{PVIopt}
\end{align}
where $\beta> 0$ is a tuning parameter.   Then $q_{\lambda^*}(\widetilde{y}|\widetilde{x})$ is the predictive
density for future data $\widetilde{y}$ given $\widetilde{x}$.   
To interpret \eqref{PVIopt}, observe that the first term, $\sum_{i=1}^n S(q_{\lambda}(\cdot|x_i),y_i)$, 
is a measure of how well the predictive densities $q_\lambda(\cdot|x_i)$ predict the observed data.  
The second term in \eqref{PVIopt} is a regularization, encouraging the pseudo-posterior
to be close to the actual posterior density.   
\cite{lai+y25} observe that $\text{KL}(q_\lambda(\theta)||p(\theta|y))$ 
can be replaced by something else, but we use the formulation in \eqref{PVIopt} here.  
As $\beta\rightarrow\infty$ the regularizer dominates the objective function, so that
optimizing \eqref{PVIopt} becomes the same as ordinary variational inference
optimizing the ELBO.  An equivalent form of the optimization \eqref{PVIopt} is
\begin{align}
 \lambda^*=&\arg \max_{\lambda} \left\{\sum_{i=1}^n S(q_\lambda(\cdot|x_i),y_i)+\beta. 
\text{ELBO}(\lambda)\right\}. \label{PVIopt2}
\end{align}
\cite{lai+y25} consider various possibilities for the scoring rule used in \eqref{PVIopt}, 
including the log score and the continuous ranked probability score.  Here we use the log score, i.e. 
$S(q_\lambda(\cdot),y_i)=\log q_\lambda(y_i)$.  

\section{Gaussian mixture PVI posteriors}

Gaussian mixture PVI posteriors for generalized linear models are 
discussed in this section.  Hierarchical extensions are developed 
later in Section 5.

\subsection{PVI for generalized linear models} \label{subsec:pvi_glm}

Consider a Gaussian mixture PVI (GM-PVI) posterior of the form
\begin{align}
 q_\lambda(\theta) & =\sum_{k=1}^K \omega_k \phi(\theta;\mu_k,\Sigma_k),   \label{GM-PVIform}
\end{align}
where $\omega_k\geq 0$ are mixing weights, $\sum_{k=1}^K \omega_k=1$, and
$\phi(\theta;\mu_k,\Sigma_k)$ is a multivariate normal density with mean $\mu_k$ and
covariance matrix $\Sigma_k$.  Write $\omega=(\omega_2,\dots, \omega_K)^\top$ with $\omega_1=1-\sum_{k=2}^K\omega_k$, 
$\mu=(\mu_1^\top,\dots, \mu_K^\top)^\top$ and $\gamma=(\text{vech}(\Sigma_1)^\top,\dots, \text{vech}(\Sigma_k)^\top)^\top$, where $\text{vech}(A)$ for a square matrix $A$ is the vector obtained
by stacking lower-triangular elements of $A$ columnwise left to right.  
The vector of variational parameters in our GM-PVI 
is $\lambda=(\omega^\top,\mu^\top,\gamma^\top)^\top$.  

It is helpful to begin with a simple case, a linear model: 
\begin{align*}
  y_i & = x_i^\top \theta+\epsilon_i,\;\;\;\epsilon_i\sim N(0,\sigma^2),
\end{align*}
where $\sigma^2$ is considered fixed for now.  
The GM-PVI posterior $q_\lambda(\theta)$ leads to 
the predictive density
\begin{align}
  q_\lambda(\widetilde{y}|\widetilde{x}) & = \int p(\widetilde{y}|\widetilde{x},\theta)q_\lambda(\theta)\,d\theta \nonumber \\
  & = \int \sum_{k=1}^K \omega_k \phi(\theta;\mu_k,\Sigma_k)\phi(\widetilde{y}|\widetilde{x}^\top \theta,\sigma^2)d\theta \nonumber \\
  & = \sum_{k=1}^K \omega_k \phi(\widetilde{y};\widetilde{x}^\top \mu_k,\widetilde{x}^\top \Sigma_k \widetilde{x}+\sigma^2).  \label{GM-PVIlin}
\end{align}
Prediction using \eqref{GM-PVIlin} is ``plug-in'' prediction for a mixture of (heteroscedastic) 
linear regression models, and
if the log score is used in PVI, the parameters $\lambda$ that are plugged in are being estimated
by fitting to the training data using regularized maximum likelihood \eqref{PVIopt2}.  
The information from the prior distribution on $\theta$ is used in the regularization term, where
$q_\lambda(\theta)$ is encouraged to be close to the posterior, which in turn depends on the prior.  
Although we assumed
$\sigma^2$ was known, this could also be estimated from the data to optimize predictive performance
along with $\lambda$.  We can also absorb $\sigma^2$ into $\theta$, but then integrating out
$\theta$ can no longer be done in closed form. An example where we do so is given in Supplementary Information~B.2. The regression components in the mixture
model are heteroscedastic linear regressions, with the form of the variance model a special
case of the approach considered in \cite{hoff+n12}.  

Next, we expand this mixture of regression interpretation of PVI to generalized linear models (GLMs).  
For the moment we consider the case where any dispersion parameters are known.  
Writing $\mu_i$ for the mean of $y_i$, it is assumed that
$g(\mu_i)=x_i^\top \theta$, for some link function $g(\cdot)$.  The sampling density for a future response is a function of 
$\theta$ only through $\widetilde{x}^\top\theta$, and we write it as $p(\widetilde{y}|\widetilde{x}^\top\theta)$.  
If we use the GM-PVI form \eqref{GM-PVIform}, 
and write the implied density of $\widetilde{x}^\top\theta$ as
$$q_\lambda(\widetilde{x}^\top\theta)=\sum_{k=1}^K \omega_k \phi(\widetilde{x}^\top\theta;\widetilde{x}^\top\mu_k,\widetilde{x}^\top \Sigma_k\widetilde{x}),$$
then the GM-PVI predictive density is
\begin{align}
  q_\lambda(\widetilde{y}|\widetilde{x}) & =\int p(\widetilde{y}|\widetilde{x}^\top \theta)q_\lambda(\widetilde{x}^\top \theta)\;d(\widetilde{x}^\top \theta) \nonumber \\
   & = \sum_{k=1}^K \omega_k \int p(\widetilde{y}|\widetilde{x}^\top\theta)\phi(\widetilde{x}^\top\theta;\widetilde{x}^\top \mu_k,\widetilde{x}^\top \Sigma_k \widetilde{x})\;d(\widetilde{x}^\top\theta). \label{GM-PVIglm}
\end{align}
Denote the integral in the $k$th term of \eqref{GM-PVIglm} as
$$I_k=\int p(\widetilde{y}|\widetilde{x}^\top\theta)\phi(\widetilde{x}^\top\theta;\widetilde{x}^\top \mu_k,\widetilde{x}^\top \Sigma_k \widetilde{x})\;d(\widetilde{x}^\top\theta).$$
If we consider a change of variable 
$$w=\frac{\widetilde{x}^\top\theta-\widetilde{x}^\top\mu_k}{\sqrt{\widetilde{x}^\top\Sigma_k\widetilde{x}}}$$
so that $\widetilde{x}^\top\theta=\widetilde{x}^\top \mu_k+w\sqrt{\widetilde{x}^\top\Sigma_k \widetilde{x}}$, 
we obtain
\begin{align*}
  I_k & = \int p\left(\widetilde{y}|\widetilde{x}^\top\mu_k+w\sqrt{\widetilde{x}^\top \Sigma_k \widetilde{x}}\right)\phi(w;0,1)\,dw,
\end{align*}
showing that the integrand is a function $f_k(w)$ multiplied by a standard normal density, so that the integral can be
approximated accurately using Gauss-Hermite quadrature 
(e.g. \citealp[Section 8.3]{sarkka+s23})
with only a small number of quadrature 
points.  Using $B$ quadrature points gives exact answers when $f_k(w)$ is any polynomial
up to degree $2B-1$.  With this approach we obtain the approximation
$$I_k\approx \sum_{b=1}^B \gamma_b.p\left(\widetilde{y}|\widetilde{x}^\top \mu_k+w_b\sqrt{\widetilde{x}^\top \Sigma_k \widetilde{x}}\right).$$
where $\gamma_b\geq 0$, $b=1,\dots, B$, are quadrature weights summing 
to $1$, and $w_b$, $b=1,\dots, B$, are the quadrature points.  
Putting this together with \eqref{GM-PVIglm}, we obtain
\begin{align}
q_\lambda(\widetilde{y}|\widetilde{x}) & \approx 
\sum_{k=1}^K \sum_{b=1}^B \omega_k \gamma_b.p\left(\widetilde{y}|\widetilde{x}^\top \mu_k+w_b \sqrt{\widetilde{x}^\top \Sigma_k \widetilde{x}}\right). \label{GM-PVIquad}
\end{align}  
The representation \eqref{GM-PVIquad} demonstrates that the GM-PVI posterior 
can be approximated by plug-in predictive inference based on a mixture of GLM's with
offset terms which are independent of the coefficients $\mu_k$.  
The PVI variational parameters $\lambda$ are estimated by
regularized maximum likelihood, similar to 
the linear model case.  Similar to linear models, 
if there are any unknown dispersion parameters, 
these could also be estimated to maximize predictive accuracy.  In the case
of the linear model, the above representation is not equivalent to the one discussed previously -- 
\eqref{GM-PVIquad} would be a mixture of homoscedastic linear models, not heteroscedastic
models as in \eqref{GM-PVIlin}.  

\subsection{PVI with covariate-dependent GM-PVI's}

Next we consider an extension where a GM-PVI is allowed to
depend on the prediction covariate, denoted $\widetilde{x}$ here, 
resulting in what we call varying GM-PVI posteriors (VGM-PVI posteriors).  
In the VGM-PVI posteriors, we allow the mixing weights in a GM-PVI posterior
to vary with $\widetilde{x}$: 
$$\mathcal{Q}=\{q_\lambda(\theta|\widetilde{x}): q_\lambda(\theta|\widetilde{x})=\sum_{k=1}^K \omega_k(\widetilde{x})\phi(\theta;\mu_k,\Sigma_k)\},$$
where 
$$\omega_k(\widetilde{x})=\frac{\exp(\widetilde{x}^\top\eta_k)}{\sum_{l=1}^K \exp(\widetilde{x}^\top \eta_l)},$$
with $\eta_1=0$ for identifiability.  

Let's consider first the linear model setting of the previous section.  
For the VGM-PVI we obtain
\begin{align}
  q_\lambda(\widetilde{y}|\widetilde{x}) & = \int p(\widetilde{y}|\widetilde{x},\theta) q_\lambda(\theta|\widetilde{x})\,d\theta \nonumber \\
  & = \int \sum_{k=1}^K w_k(\widetilde{x})\phi(\theta;\mu_k,\Sigma_k)\phi(\widetilde{y}|\widetilde{x}^\top \theta,\sigma^2)\,d\theta \nonumber \\
  & = \sum_{k=1}^K \omega_k(\widetilde{x}) \phi(\widetilde{y};\widetilde{x}^\top \mu_k,\widetilde{x}^\top \Sigma_k \widetilde{x}+\sigma^2).  \label{VGM-PVIlin}
\end{align}
This is similar to \eqref{GM-PVIlin}, with the mixing weights changing from constant to covariate
dependent.  Similarly in the GLM case we obtain analogous to \eqref{GM-PVIquad}
\begin{align}
q_\lambda(\widetilde{y}|\widetilde{x}^\top\theta) & \approx 
\sum_{k=1}^K \sum_{b=1}^B \omega_k(\widetilde{x}) \gamma_b.p(\widetilde{y}|\widetilde{x}^\top \mu_k+w_b \sqrt{\widetilde{x}^\top \Sigma_k \widetilde{x}}). \label{VGM-PVIquad}
\end{align}
So again we obtain some equivalences between VGM-PVI posteriors and plug-in prediction for mixture of experts models, but now with covariate-dependent mixture 
weights. 

An additional aspect of VGM-PVI posteriors is that we need to decide what the regularization term in the PVI objective function should be.  Here we suggest to use the KL divergence between 
$$\overline{q}_{\lambda}(\theta)=\frac{1}{n}\sum_{i=1}^n q_\lambda(\theta|x_i)=\sum_{k=1}^K \Bar{\omega}_k\phi(\theta;\mu_k,\Sigma_k)$$
and $p(\theta|y)$.  The KL divergence term in the regularizer becomes equivalent to the 
ordinary ELBO for the density $\overline{q}_{\lambda}(\theta)$ for optimization purposes. The density $\overline{q}_{\lambda}(\theta)$ is a Gaussian mixture, where the mixing weights, $\Bar{\omega}_k=n^{-1}\sum_{i=1}^n\omega_k(x_i)$, are the average of the mixing weights in the $q_\lambda(\theta|x_i)$. However, there are many alternative ways we can obtain a set of average PVI posteriors over covariates
close to the true posterior.  
The regularizer becomes
\begin{align*}
    &\sum_{k=1}^K\overline{\omega}_k\mathbb{E}_{\phi(\theta;\mu_k,\Sigma_k)}[\log p(y\mid\theta) + \log p(\theta)]+\mathbb{E}_{\overline{q}_{\lambda}(\theta)}[-\log\overline{q}_{\lambda}(\theta)]\\
    &=\sum_{k=1}^K n^{-1}\sum_{i=1}^n \mathbb{E}_{\phi(\theta;\mu_k,\Sigma_k)}[\omega_k(x_i) \log\left(p(y_i\mid\theta)p(\theta)\right)]+\mathbb{E}_{\overline{q}_{\lambda}(\theta)}[-\log \overline{q}_{\lambda}(\theta)].
\end{align*}
So the regularizer encourages the ``average PVI posterior'' to be close to the posterior. In particular, the first term forces each mixture component to be close to the weighted posterior for the sub-data in the region on which the component is actually active. For the expected log-likelihood terms, expectations are only taken over single Gaussians, so that these terms can be derived in closed form for many common GLMs. Alternatively, similar to the computation of the predictive densities $q_\lambda(y_i|x_i)$, Gauss-Hermite quadrature can be employed. The second term acts as a global penalty for the complexity of the mixture model in terms of it's entropy. This depends only on the Gaussian mixture density $\overline{q}(\theta)$, and we use a closed-form approximation for entropy of Gaussian mixtures given in \citet{huber+bdh08}. We give analytic expressions for the optimization targets for all examples considered within the manuscript in Supplementary Information~A. 

To ensure positive definiteness of the mixture covariance matrices, $\Sigma_k$, we consider the Cholesky decomposition $\Sigma_k=C_kC_k^\top$ with a lower triangular matrix $C_k$ with positive diagonal. Writing $A^*$ for any matrix $A$, such that $A^*_{ij}=A_{ij}$ if $i\not=j$ and $A_{jj}^*=\log A_{jj}$, the non-zero entries of $C_k^*$ are unrestricted. Writing $\eta=(\eta_2^\top,\dots,\eta_K^\top)^\top$, $\mu=(\mu_1^\top,\dots,\mu_K^\top)^\top$, and $c=(\text{vech}(C_1^*)^\top,\dots,\text{vech}(C_K^*)^\top)^\top$, the vector of variational parameters in VGM-PVI is $\lambda=(\eta^\top,\mu^\top,c^\top)^\top$. Due to the closed form of the optimization target, $\lambda$ is optimized using a simple gradient ascent (GA) scheme with the adaptive Adam learning rate \citep{KinBa2014}. 

A particular challenge when working with mixture models is selecting the number of mixture components. In the examples discussed below the number of mixture components is chosen along the following heuristic. We initialize GA with a large number of potential components set to $K=10$ for the GLM examples, and to $K=5$ in all other examples. After $2,000$ steps of GA all components $k$ for which $\omega_k(x_i)<\omega_l(x_i)$ for all $i=1,\dots,n$, $l\not=k$, are removed from the model. This procedure is repeated until no further components are removed and the algorithm is then run until convergence. Even though there is no theoretical guarantee that this heuristic results in an optimal choice for $K$, we found this to be a fast way of selecting a reasonable model in all examples considered. Alternatively, models for several choices of $K$ can be trained and the best one can then be selected based on some model selection criterion.
The recent work by \cite{davies+mos25} could also be adapted to perform predictive variational model averaging.

\section{Experiments and Applications}\label{sec:GLM_Examples}
This section illustrates VGM-PVI in several GLM tasks. An additional simulation study and an example with a Gaussian likelihood model are given in Supporting Information~B.1 and B.2 respectively.

\subsection{Simulations -- Logistic classification}\label{subsec:sim_logistic}
As a first example, we consider a synthetic binary classification task from \citet{MclCheFraKno2025}. For $i=1,\dots,n$, covariates $x_i$ are uniformly drawn on $[-2,2]^2$, and labels $y_i$ are deterministically assigned to data in the top-left and the bottom-right quadrant, while labels in the remaining two quadrants are assigned randomly. That is
\begin{align*}
    y_i\vert x_i \sim \begin{cases}
        \BernD(0) \text{ if } x_{i1}<0, x_{i2}>0;\\
        \BernD(1) \text{ if } x_{i1}>0, x_{i2}<0;\\
        \BernD(0.5) \text{ else},
    \end{cases}
\end{align*}
where $\BernD(p)$ denotes the Bernoulli distribution with success probability $p$.
We train a logistic regression model $y_i\sim\BernD(\sigma(x_i^\top\theta))$ with prior $\theta\sim\ND(0,2.5^2I_2)$, where $\sigma(\cdot)$ denotes the logistic function, to a data set with $n=1,000$ observations. Predictions based on the logistic regression model are incapable of recovering the non-linearities introduces by the true data generating process (DGP). We consider the average log point-wise predictive density (llpd),$$\text{llpd}=\frac{1}{n_\text{test}}\sum_{i=1}^{n_\text{test}}\log \int p(y^{(\text{test})}_i\mid x^{(\text{test})}_i, \theta)q_\lambda(\theta)d\theta=\frac{1}{n_\text{test}}\sum_{i=1}^{n_\text{test}}S(q_\lambda(\cdot),y^{(\text{test})}_i),$$ evaluated on a large hold-out test set with $n_\text{test}=100,000$ observations as a performance metric. Larger values indicate better predictions.  

\begin{figure}[tbh]
    \centering
    \includegraphics[width=0.95\linewidth,keepaspectratio]{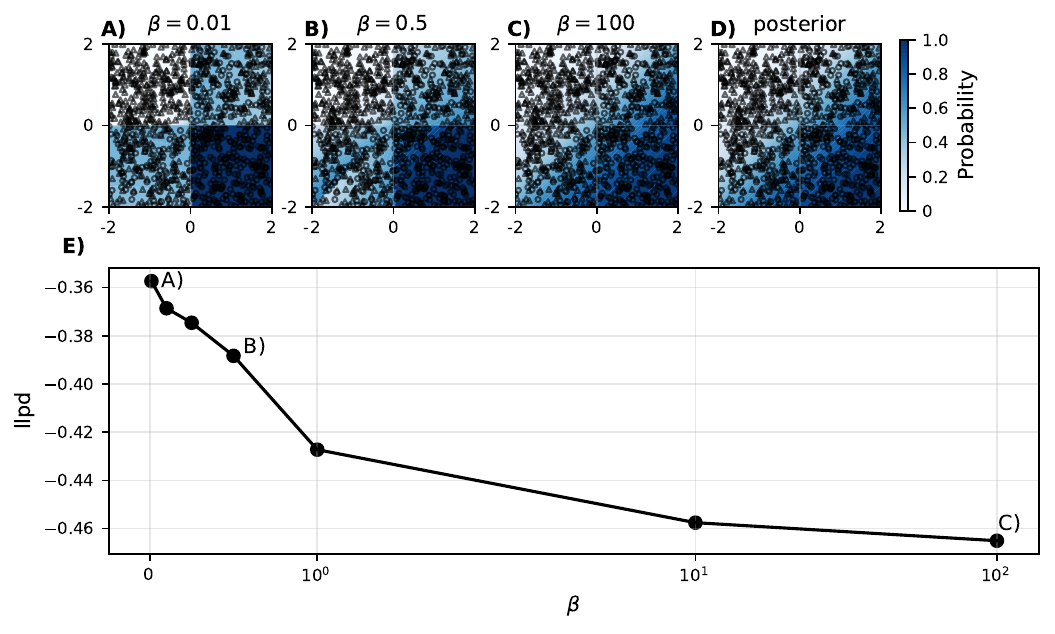}
    \caption{\small Simulation -- Logistic regression. Posterior predictive distribution for different for \textbf{A)} $\beta=0.01$, \textbf{B)} $\beta=0.5$, \textbf{C)} $\beta=100$, and \textbf{D)} the true posterior. The observed data is given as a scatter plot, where triangles denote $y=0$ and circles $y=1$. \textbf{E)} plots the llpd on the hold-out test data versus different values of $\beta$.}
    \label{fig:sim_binomial} 
\end{figure}

Figure~\ref{fig:sim_binomial} compares the performance of VGM-PVI for several choices of the hyperparameter $\beta$. The first row summarizes the posterior predictive distributions. For large $\beta$ VGM-PVI is similar to the true posterior predictive distribution, while $\beta=0.01$ matches the true DGP closely. Varying $\beta$ allows for a smooth interpolation between these two extremes. Figure~\ref{fig:sim_binomial} E) shows the llpd evaluated on the hold-out test set. The largest llpd-value is achieved by $\beta=0.01$. Under this specification, VGM-PVI selects $k=4$ mixture components using the heuristic described above matching the four disjoint regions present in the DGP.

\subsection{Logistic regression --  Gamma telescope data}
We consider a challenging classification task from astrophysics using Monte Carlo data for an imaging gamma-ray Cherenkov telescope \citep{Boc2004}. The goal is to discriminate observations caused by incident gamma rays (signal) from hadronic showers (noise). The data set consists of $n=19,020$ observations on $d=10$ variables summarizing the Cherenkov images. A full description of the data is given in \citet{BocChiGauWit2004}. As in \citet{BocChiGauWit2004}, we use a randomly sampled subset of $2/3$ of the data for training and the remaining data for evaluation. The base model is a logistic regression model, where the log-odds are linked to an intercept and linear effects for all covariables. After standardisation we apply the prior $\theta\sim\ND(0,2.5^2I_{11})$. 

The receiver operating characteristic (ROC) curve on the hold-out test data is shown in Supplementary Information~B.3. The ROC curve displays the true positive rate (TPR) as a function of the false positive rate (FPR) with primary gamma rays considered to be the positive class. In practice, classifying a background event as signal is worse than classifying a signal event as a hadronic shower, and therefore we are in particular interested in the TPR when the FPR is low. Table~\ref{tab:gamma} reports the TPR for a number of low thresholds for the FPR. For large values of $\beta$, VGM-PVI performs similar to the exact posterior predictive distribution. Decreasing $\beta$, however, results in a drastic improvement for out-of-sample classification for all thresholds for the FPR. VGM-PVI with a small penalty, $\beta=0.01$, increases the TPR by a factor of $4$ compared to the conventional posterior predictive conditional on $\text{FPR}=0.01$. Earlier analyses of this data set found that classification trees and Bayes classifiers based on sophisticated density estimators are among the best performing classifiers \citep{BocChiGauWit2004,NagCza2016}. These methods can relate the class probabilities to complex interactions in the $10$-dimensional covariate space, while a simple logistic regression model without interaction terms is incapable of modelling cross-feature dependencies. The predictive distribution resulting from VGM-PVI, \eqref{VGM-PVIquad}, however, is a mixture of logistic regression models with covariate dependent weights and can thus capture some dependencies between features. This is likely the reason for the drastic increase in performance from VGM-PVI compared to the base model.

\begin{table}[tbh]
    \centering
    \begin{tabular}{l|ccccc}
        FPR & 0.01 & 0.02 & 0.05 & 0.1 & 0.2 \\
        \hline\hline
        VGM-PVI ($\beta=0.01$) &\textbf{0.3185} & \textbf{0.4029} & \textbf{0.5787} & \textbf{0.7396} & \textbf{0.8940} \\
        VGM-PVI ($\beta=0.5$) & 0.2520 & 0.3549 & 0.5518 & 0.6993 & 0.8422\\
        VGM-PVI ($\beta=1.0)$ & 0.2043 & 0.3182 & 0.5161 & 0.6964 & 0.8501\\
        VGM-PVI ($\beta=100$) & 0.0827 & 0.1717 & 0.3295 & 0.5173 & 0.7350\\
        Posterior predictive & 0.0775 & 0.1700 & 0.3194 & 0.5115 & 0.7271\\
        \hline\hline
    \end{tabular}
    \caption{\small Gamma telescope. Comparison of TPR for a given target FPR for the hold-out test set data under both the conventional posterior predictive for the base model and different specifications of VGM-PVI with varying $\beta$. Bold values mark the best method per column. }
    \label{tab:gamma}
\end{table}

\subsection{Poisson likelihood model -- AIDS case counts}
To further illustrate how VGM-PVI can be helpful for model criticism and model improvement, we reanalyse quarterly counts of reported AIDS cases in the United Kingdom (UK) from January 1983 to September 1990 \citep{StaRig1992} using a Poisson GLM. The six-dimensional linear predictor consists of an intercept, a linear and a quadratic term in time, as well as binary dummy variables to capture quarterly seasonality. 
Due to the relatively low number of observations in this data set, a proper test-training split is infeasible. We consider therefore the widely applicable information criterion (WAIC) suggested in \citet{VehGelGab2017} as a model selection criterion to determine $\beta$. The optimal penalty parameter $\beta^*$ is then given as
\begin{equation}\label{eq:elpd}
\beta^*=\arg\max_\beta\left\{\sum_{i=1}^n\log\left(\frac{1}{M}\sum_{m=1}^Mp(y_i\mid\theta_m)\right)-\sum_{i=1}^n V_{m=1}^M \left(\log(p(y_i\mid\theta_m))\right)\right\},
    \end{equation}
where $V_{m=1}^M(\cdot)$ is the sample variance and $\theta_1,\dots,\theta_M\sim q_{\lambda}(\theta)$ is a sample from the variational posterior for a fixed $\beta$. We solve optimization problem \eqref{eq:elpd} using Bayesian optimization \citep[BO; e.g.,~][]{ShaSweWanAdaDef2016} exploring values between $0.01$ and $100$. BO efficiently explores the candidate space by building a probabilistic model of the objective \eqref{eq:elpd} as a function of $\beta$, making it a computationally attractive alternative to simple grid search.

\begin{figure}[tbh]
    \centering
    \includegraphics[width=0.95\linewidth,keepaspectratio]{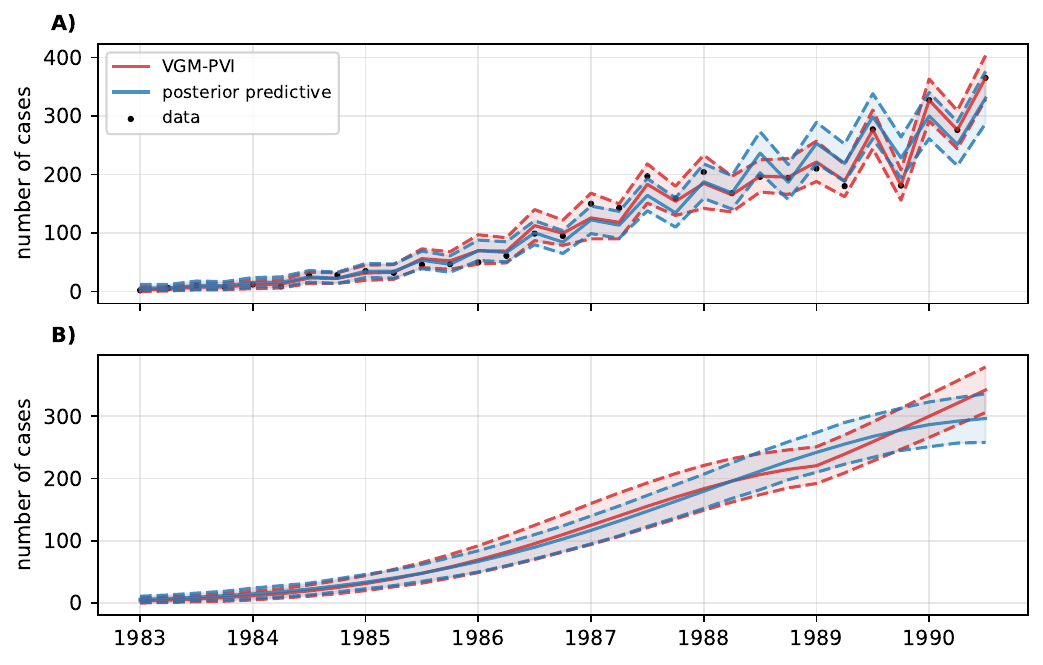}
    \caption{\small AIDS case counts. Panel \textbf{A)} shows the mean (bold) and a 95\% credible interval (shaded) derived by VGM-PVI (red) and the true posterior (blue) as well as the observed data (points).  Panel \textbf{B)} shows the deseasonalized temporal trend for both methods.}
    \label{fig:aids} 
\end{figure}

Figure~\ref{fig:aids} A) compares the predictive distribution under the conventional posterior with that under VGM-PVI against the observed case counts. Even though a visual inspection suggests that the base model captures the data reasonably well, a small value for the penalty parameter, $\beta\approx 0.01$, was selected via \eqref{eq:elpd} indicating that the model extension provided by VGM-PVI improves the base model. While both methods detect a similar seasonality with higher counts in the first and third quarters than in the second and fourth, the estimated temporal trends differ. Figure~\ref{fig:aids} B) shows the deseasonalised time series obtained by fixing the quarterly dummy variables at equal values. The base model yields an S-shaped temporal trend, while VGM-PVI detects an initial period of quadratic growth followed by a flattening around the beginning of 1988 before a renewed increase towards the end of the observation period. This more nuanced temporal trend matches findings in previous studies. \citet{StaRig1992} postulate that the introduction of antibody tests in the UK resulted in a change of behaviour for high-risk groups and consequently to a breakpoint in the temporal trend around July 1987. This dynamic cannot be captured by the linear predictor specified in the base model, a potential flaw that can be discovered by comparing the conventional posterior predictive to VGM-PVI.

\section{PVI for hierarchical models}
To illustrate how VGM-PVI can be extended to hierarchical models we analyse data on water temperature measured at a monitoring station at Upper Peirce reservoir in Singapore \citep{TanNot2014}. The data are measured on $n=290$ days during the period September 9, 2010 to August 10, 2011 at $g=11$ different water depths from the water surface: 0.5m, 2m, 4m, 6m, 8m, 10m, 12m, 14m, 16m, 18m, and at the bottom. Writing $y_{ij}$ for the temperature on day $t_i$, $i=1,\dots,n$, at depth $j=1,\dots,g$, we consider a model of the form 
\begin{equation}\label{eq:temperature_model}
    y_{ij}=f(t_i)+a_i+b_j+\varepsilon_{ij},
\end{equation}
where $a_i\sim\ND\left(0,\sigma^2_a\right)$ is an observation specific random effect, $b_j\sim\ND\left(0,\sigma^2_b\right)$ is a random intercept for depth, and $\varepsilon_{ij}\sim\ND(0,\sigma^2_\varepsilon)$ represents measurement error variance. We will assume that $\sigma^2_a, \sigma^2_b$, and $\sigma^2_\varepsilon$ are known. After transforming $t\in[0,1]$, we model the non-linear temporal effect as $f(t)=\beta_0+\beta_1t+\beta_2t^2+\beta_3t^3=x^\top\beta$ with $x=(1,t,t^2,t^3)^\top$ and assume a weak prior $\beta=\left(\beta_0,\beta_1,\beta_2,\beta_3\right)^\top\sim\ND(0,100^2)$. Writing $b=(b_1,\dots,b_g)^\top$ and $a=(a_1,\dots,a_n)^\top$, the stacked vector of unknown model parameters is $\theta=(\beta^\top,a^\top,b^\top)^\top$. Our interest lies in predicting a vector of water temperatures $\Tilde{y}=\left(\Tilde{y}_1,\dots,\Tilde{y}_g\right)^\top$ at time $\Tilde{t}$ at the previously observed water depths. For this reason, we consider a VGM-PVI of the form 
\begin{equation*}
    q_\lambda(\theta\vert x)=\left[\sum_{k=1}^K\omega_k(x) \phi(\theta_{\backslash a};\mu_k,\Sigma_k)\right]\prod_{i=1}^n \phi(a_i;m_i,\tau_i^2),
\end{equation*}
where $\theta_{\backslash a}$ denotes the vector of model parameters excluding the vector of observation specific intercepts $a$, and $m_i$, $\tau_i^2$, $i=1,\dots,n$ are additional variational parameters to be learned with $\lambda$. Then,
\begin{align*}
    q_\lambda(\Tilde{y}\mid \Tilde{x})&=\int \omega_k(\tilde{x}) \phi(\Tilde{y};\Tilde{X}\theta_{\backslash a},(\sigma^2_\varepsilon+\sigma^2_a)I_g) \phi(\theta_{\backslash a};\mu_k,\Sigma_k)d\theta_{\backslash a}\\
    &=\sum_{k=1}^K \omega_k(\tilde{x}) \phi\left(\Tilde{y};\Tilde{X}^\top\mu_k,\Tilde{X}^\top\Sigma_k\Tilde{X}+(\sigma^2_\varepsilon+\sigma^2_a)I_g\right),
\end{align*}
where $\Tilde{X}$ is a stacked design matrix matching the model \eqref{eq:temperature_model}. Again, the optimal penalty parameter $\beta^*$ is selected by optimizing the WAIC in \eqref{eq:elpd}. We select $\beta^*=0.01$.

\begin{figure}
    \centering
    \includegraphics[width=0.9\linewidth]{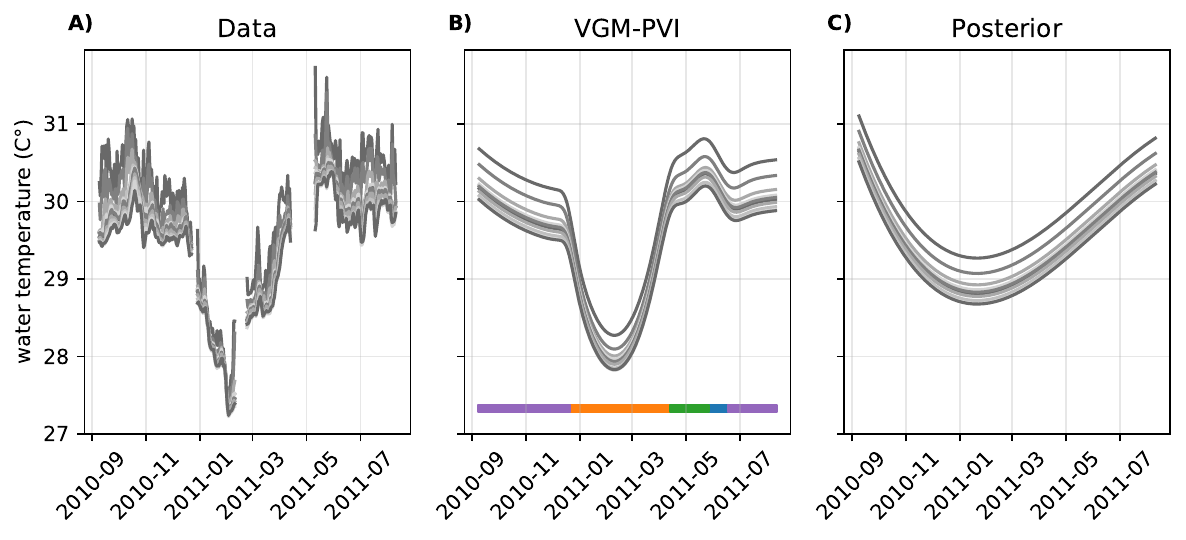}
    \caption{\small Temperature data. Water temperature over time at different depths (indicated by gray shades). Shown is \textbf{A)} the observed data, as well as the estimated mean functions $f(t)+b_j$ for the different depths under \textbf{B)} VGM-PVI and \textbf{C)} the conventional posterior. \textbf{B)} also shows the dominating mixture component under VGM-PVI at each point in time (colour bar). Different clusters are indicated by colour.}
    \label{fig:temperature}
\end{figure}

Figure~\ref{fig:temperature} compares the observed data with the estimated mean functions $f(t)+b_j$ for the different depths, $j=1,\dots,g$, over time under VGM-PVI and the conventional posterior. The estimated mean functions have vastly different shapes. VGM-PVI results in a sharp decline between January and April that is also apparent in the data, but cannot be fully captured by the cubic polynomial the base model is restricted to. At irregular points in time aeration devices were operated to mix the water at Upper Peirce Reservoir, and it is expected that the temperature is more uniform with depth at those times. Under the base model, however, functions for different depths can only differ by a random intercept. VGM-PVI allows for more flexibility and the plot shows a clear temporal effect on the expected differences of temperature at different depths so that at least some of the heterogeneous temperature layering can be captured. 

In this example, the covariate dependent mixture weights for VGM-PVI are effects of time only. VGM-PVI results therefore in an implicit clustering of time-periods and Figure~\ref{fig:temperature} B) visualizes this clustering by showing $\arg\max_k \omega_k(t)$ as a function of time. VGM-PVI selects four clusters corresponding to distinguished periods. The largest (purple) cluster corresponds to the two periods September 2010 to December 2010 and June 2011 to August 2011. In these periods VGM-PVI is closest to the conventional posterior. The second largest cluster (orange) corresponds to the period January 2011 to April 2011. In this period the water temperature plunges at all depths and VGM-PVI differs most markedly from the conventional posterior. 

\section{PVI for latent Gaussian process models}\label{sec:PVI_gp}
Lastly, we consider the scenario where the likelihood term for the response $y_i$  takes the form $p(y_i|f(x_i))$, depending on a latent function $f(\cdot)$ evaluated at the covariates $x_i$. The latent function is given a zero mean Gaussian process prior with covariance function $\delta(\cdot,\cdot)$. We write $f=(f(x_1),\dots, f(x_n))^\top$. \cite{dezfouli+b15} have considered variational inference with Gaussian mixture families for these latent GP models, and we adapt their approach here to the case of VGM-PVI. See also \cite{bonilla+kd19,fan+bos23} for a more detailed discussion of the approach and some extensions. Their approach is highly scalable, and uses inducing point approximations;  see, 
for example, \cite{quinonero-candela+r05} and \cite{titsias09}.  Inducing point methods for GPs are related to a variety of scalable basis function approaches used 
in spatial statistics \citep{banerjee+gfs08,cressie+sz22}. 

Inducing point approximations of GPs consider an augmented posterior distribution where values of the latent process (inducing values) are considered at inducing points (covariates) denoted here by $z_{1},\dots, z_{m}\in \mathcal{X}$. The introduction of the inducing points allows computations to be done using matrices of size $m$ rather than $n$, where $m\ll n$ in the case where $n$ is large. We define $\breve{f}=(f(z_1),\dots, f(z_m))^\top$. Henceforth we write $X$ for the matrix with rows $x_1,\dots, x_n$, and $Z$ for the matrix with rows $z_{1},\dots, z_{m}$. For the posterior
 $$p(f,\breve{f}|y)\propto p(\breve{f})p(f|\breve{f})p(y|f)$$
 an approximating family of the form 
 \begin{align}
  q_\lambda(f,\breve{f}) & =q_\lambda(\breve{f})p(f|\breve{f}),  \label{gpapprox}
 \end{align}
 is used, in which the conditional prior is used as the 
 conditional variational posterior for $f|\breve{f}$.  
 By standard properties of multivariate normal distributions, the conditional prior can be
 expressed
 \begin{align}
   p(f|\breve{f}) = \phi(f;c,C),  \label{condprior}
 \end{align}
 where defining $A:=\delta_j(X,Z)\delta(Z,Z)^{-1}$, 
 $c=A\breve{f}$ and $C=\delta(X,X)-A\delta(Z,X)$. Given two matrices $X'$ and $X''$ with rows in $\mathcal{X}$,  $\delta(X', X'')$ is the matrix whose $(k, l)$-th entry is $\delta(\cdot,\cdot)$ evaluated for the $k$-th row of $X'$ and the $l$-th row of $X''$. 
 For the marginal variational posterior of inducing variables, a Gaussian mixture is assumed,
 \begin{align}
   q_\lambda(\breve{f}) & = \sum_{k=1}^K \omega_k \phi(\breve{f};\mu_k,\Sigma_k), \label{indmarg}
 \end{align}
 where we assume here that $\Sigma_k$ is diagonal.  
 From \eqref{condprior} and \eqref{indmarg}, we can obtain the marginal for $f$ in \eqref{gpapprox},
 $$q_\lambda(f)=\sum_{k=1}^K \omega_k \phi(f;\alpha_{k},\Omega_{k}),$$
 where $\alpha_{k}=A\mu_{k}$ and $\Omega_{k}=C-A\Sigma_{k}A^\top$.  
 In computing the ELBO and the gradient of the ELBO, we can easily obtain the marginal 
 for $f(x_i)$ in the above expression, and computing required expectations only requires
 expectations with respect to univariate marginal distributions. 

To illustrate VGM-PVI for latent GP models, we consider the popular LIDAR data benchmark \citep{HolHoeBjoRagEdn1996,RupWanCar2003}. The data set describes the reflection of laser-emitted light to detect chemical compounds in the atmosphere. The response variable, $y$, is the logarithm of the ratio of received light from two laser sources. We consider the distance travelled before the light is reflected back to its source as the only covariate $x$ and a Gaussian likelihood for $y_i$, $\phi(y_i;f(x_i),\sigma^2)$. We standardize both the response and the regressor variable.

For the moment, we assume that the hyperparameter $\sigma^2$ is fixed, and we discuss its estimation later. In this case, the predictive distribution under VGM-PVI for an unobserved $\widetilde{y}$ for observed covariate $\widetilde{x}$ is
 \begin{align*}
   q_\lambda(\widetilde{y}|\widetilde{x}) & = \sum_{k=1}^K \omega_k(\widetilde{x}) \phi(\widetilde{y};\mu(\widetilde{x}),
   \sigma^2(\widetilde{x})+\sigma^2),
 \end{align*}
 where 
 \begin{align*}
   \mu(\widetilde{x}) & = \delta(\widetilde{x},Z)\delta(Z,Z)^{-1}\mu_k \\
   \sigma^2(\widetilde{x}) & = \delta(\widetilde{x},\widetilde{x})-\delta(\widetilde{x},Z)\left\{
   \delta(Z,Z)^{-1}-\delta(Z,Z)^{-1}\Sigma_k\delta(Z,Z)^{-1} \right\}\delta(Z,\widetilde{x}).
 \end{align*}
 In PVI with the log score and the above predictive densities, prediction corresponds 
 to plug-in prediction after fitting a mixture of experts of heteroscedastic linear models -- the mean in 
 the $k$th component involves predictors given by the basis functions in $\delta(Z,Z)^{-1}\delta(Z,\widetilde{x})$, with coefficients $\mu_k$. For a 
 non-Gaussian likelihood, we could use Gauss-Hermite quadrature similar to the GLM case to approximate the predictive distributions, obtaining a mixture of experts of GLMs.
 
A particular challenge in the LIDAR data is the pronounced heteroscedasticity, which makes it difficult to set $\sigma^2$ upfront. \citet{NguBon2014} suggest to learn $\sigma^2$ jointly with the variational parameters by optimizing the ELBO, and we extend this idea to VGM-PVI. To do so, we augment a component specific Dirac measure for $\sigma^2$ independent of $f$ into each mixture component of the variational approximation. This way, $\sigma^2$ can vary with the covariates $x$ through the covariate dependent weights $\omega_k(x)$. While \citet{NguBon2014} also learn the hyperparameters of the Gaussian process prior from the data, we consider them fixed. 

\begin{figure}[tbh]
    \centering
    \includegraphics[width=0.9\linewidth]{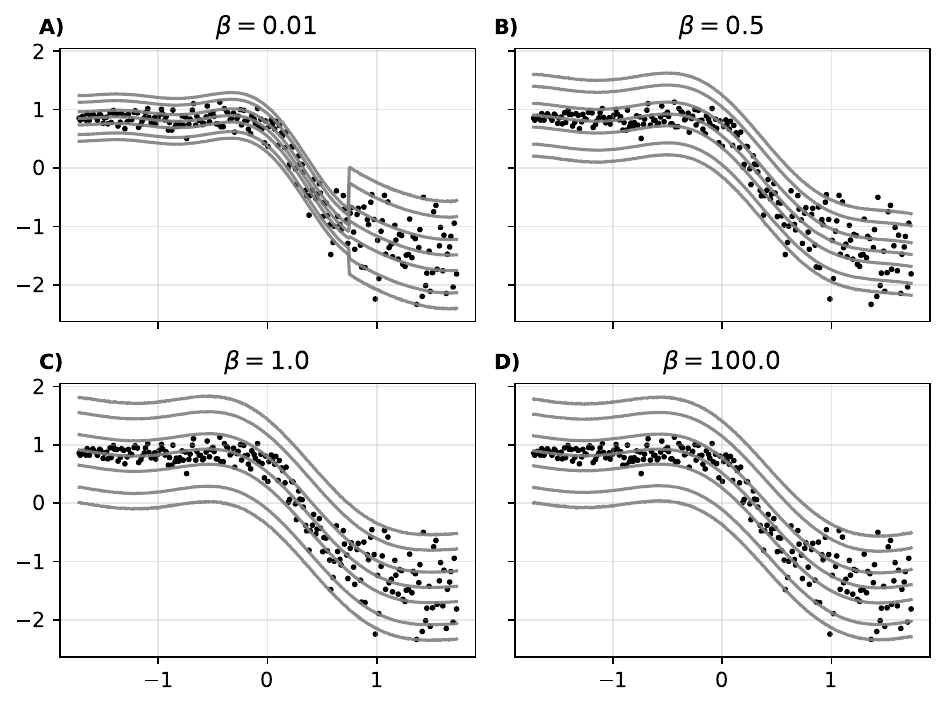}
    \caption{\small Lidar data. Posterior predictive distributions for  \textbf{A)} $\beta=0.01$, \textbf{B)} $\beta=0.5$, \textbf{C)} $\beta=1.0$, and \textbf{D)} $\beta=100$. Gray lines correspond to quantiles at levels $\alpha=0.01,0.05,0.25,0.5, 0.75, 0.95, 0.99$, and the points indicate the observed data.}
    \label{fig:lidar}
\end{figure}

The Lidar data has $n=221$ observations, so that we consider $m=n$ in this application. Figure~\ref{fig:lidar} plots quantile lines for the estimated predictive distribution under VGM-PVI for several choices of $\beta$ versus the observations. For all choices of $\beta$, the non-linear mean function matches the data well. However, the base model cannot capture the heteroscedasticity in the data and thus for large $\beta$ uncertainty is overestimated for small $x$, while smaller values of $\beta$ capture the varying variance. For $\beta=0.01$, the variance does not change gradually and an abrupt jump in uncertainty is clearly visible. This abrupt transition is inherent to VGM-PVI as the covariate dependent mixture weights are only indirectly penalized towards smoothness in $x$ through regularization towards the conventional posterior, and thus VGM-PVI results in a more gradual transition as $\beta$ increases. A non-gradual transisiton, however, can be helpful for diagnostics as it points towards regions where the base model might be misspecified.

\section{Discussion}\label{sec:Discussion}
We have demonstrated that PVI can be used not only to improve predictive performance under misspecification, but also to provide interpretable diagnostics that guide principled model expansions. By considering structured regression settings, we can obtain predictive distributions under GM-PVI via one-dimensional quadrature. The resulting predictive mixtures coincide with mixture-of-experts expansions of the baseline model, first with covariate-independent weights and then, under VGM-PVI with covariate-dependent weights. The components and weights of the predictive mixture indicate where the baseline parameterization is inadequate and how parameters should vary across the covariate space to recover improved prediction in a conventional Bayesian framework. 

Good performance was demonstrated in applications to linear, logistic, and Poisson regression, as well as hierarchical models with random effects, and latent Gaussian process models. In these examples, the covariate dependent weights of VGM-PVI correspond to gating functions in mixture-of-experts models, highlighting interactions, non-linearities, or non-stationarity in latent processes that the baseline model fails to capture. These factors act as diagnostic signals indicating where model expansions are most likely to be beneficial. 

While the Gaussian mixture structure is highly expressive, it comes with the usual computational challenges inherent to mixture modelling: Mixture components are only identified up to permutation, and selecting an optimal number of mixture components can be challenging. Here, we proposed a simple heuristic to select a reasonable number of mixture components in a one-shot procedure jointly with learning the variational parameters. While this worked well in the examples considered, external model selection criteria might become necessary in more complex base models \cite[e.g.][]{davies+mos25}. Our approach considered the log-score and a penalization towards the conventional posterior, but depending on the underlying task alternative scoring rules and penalizations could be explored. One merit of the log-score is that the optimization target can be derived in closed form, which allows for direct optimization of the variational parameters. Many alternative scoring rules can be written as an expectation over the variational distribution. In this case, stochastic gradient ascent, where a version of the reparametrization trick \citep{KinWel2014,RezMohWie2014,xu+qrs19} approximates the gradient using a sample from $q_\lambda(\theta)$, can be used for training. 

The local dependence on the covariates through the mixture weights in VGM-PVI departs from conventional Bayesian predictive inference, and a better theoretical understanding of this and the effects of different regularizers in the PVI optimization is an important direction for future research. 
In applications where PVI is used as a tool for model criticism, it is important
that the PVI posterior has a structured form that is interpretable in terms
of model expansions;  although we have used Gaussian mixtures in this work, 
there are 
other possibilities, and what is most insightful may depend on the statistical
model used.

%% file: appendix.tex
\section{Analytic expression of optimization targets} \label{app:targets}
This section gives analytic expressions for the optimization targets for the examples considered in the manuscript. The target of the optimization problem is
\begin{align*}
    &\sum_{i=1}^n S(q_\lambda(\cdot),y_i)+\beta.\left(\mathbb{E}_{\overline{q}_{\lambda}(\theta)}[\log p(y\mid\theta)p(\theta)]+\mathbb{E}_{\overline{q}_{\lambda}(\theta)}[-\log\overline{q}_{\lambda}(\theta)]\right)\\
    =&\sum_{i=1}^n S(q_\lambda(\cdot),y_i)+\beta.\left(\sum_{k=1}^K\overline{\omega}_k\left\{\mathbb{E}_{\phi(\theta,\mu_k,\Sigma_k)}[\log p(y\mid\theta)]+\mathbb{E}_{\phi(\theta,\mu_k,\Sigma_k)}[\log p(\theta)]\right\}+\mathbb{E}_{\overline{q}_{\lambda}(\theta)}[-\log\overline{q}_{\lambda}(\theta)]\right).
\end{align*}

Approximations to the score term, $S(q_\lambda(\cdot),y_i)$ are discussed in the manuscript directly. The entropy term, $\mathbb{E}_{\overline{q}_{\lambda}(\theta)}[-\log \overline{q}_{\lambda}(\theta)]$ is approximated by a lower bound  due to \citet{huber+bdh08} in all examples
\begin{align*}
    \mathbb{E}_{\overline{q}_{\lambda}(\theta)}[-\log\overline{q}_{\lambda}(\theta)]\approx -\sum_{k=1}^K \Bar{\omega}_k \log\left(\sum_{l=1}^K\Bar{\omega}_l\phi(\mu_k;\mu_l;\Sigma_k+\Sigma_l)\right).
\end{align*}
Terms for the expected log-likelihood, $\mathbb{E}_{\phi(\theta,\mu,\Sigma)}[\log p(y\mid\theta)]$, and for the expected log-prior $\mathbb{E}_{\phi(\theta,\mu,\Sigma)}[\log p(\theta)]$ are model specific and we give closed form derivations for the examples considered in the manuscript below. Note that the expectation is only taken over a single mixture component $\phi(\cdot;\mu,\Sigma)$. 

\subsection{Expected log-prior}
\paragraph{Independent Gaussian prior} Assume $\theta\sim\ND(0,\tau^2I_p)$ with $\tau>0$ fixed. Then, 
\begin{align*}
    \mathbb{E}_{\phi(\theta,\mu,\Sigma)}[\log p(\theta)]&=\mathbb{E}_{\phi(\theta,\mu,\Sigma)}\left[-\frac{p}{2}\log(2\pi\tau^2)-\frac{1}{2\tau^2}\theta^\top\theta\right] \nonumber \\ 
    &= -\frac{p}{2}\log(2\pi\tau^2)-\frac{1}{2\tau^2}\left\{\mu^\top\mu+\trace\Sigma\right\}. 
\end{align*}

\paragraph{Gaussian prior} Let $\theta\sim\ND(0,\Omega)$ for a known covariance matrix $\Omega$. Then,
\begin{align*}
    \mathbb{E}_{\phi(\theta,\mu,\Sigma)}[\log p(\theta)]&=\mathbb{E}_{\phi(\theta,\mu,\Sigma)}\left[-\frac{p}{2}\log(2\pi)-\frac{1}{2}\log\det\Omega-\frac{1}{2}\theta^\top\Omega^{-1}\theta\right]\\
    &=-\frac{p}{2}\log(2\pi)-\frac{1}{2}\log\det\Omega-\frac{1}{2}\mu^\top\Omega^{-1}\mu+\trace(\Omega^{-1}\Sigma).
\end{align*}

\subsection{Expected log-likelihood}
\paragraph{Gaussian likelihood with fixed variance}
Consider the model $y_i\sim\ND(x_i^\top\theta,\sigma^2)$ with $\sigma^2>0$ fixed. Then,
\begin{align*}
    \mathbb{E}_{\phi(\theta,\mu,\Sigma)}[\log p(y\mid\theta,x)] &= \sum_{i=1}^n\mathbb{E}_{\phi(\theta,\mu,\Sigma)}\left[-\frac{1}{2}\log(2\pi\sigma^2)-\frac{1}{2\sigma^2}(y_i-x_i^\top\theta)^2\right]\\
    &=-\frac{n}{2}\log(2\pi\sigma^2)-\frac{1}{2\sigma^2}\sum_{i=1}^n\left\{(y_i-x_i^\top\mu)^2+x_i^\top\Sigma x_i\right\}
\end{align*}

\paragraph{Linear regression with unknown variance}
We consider $y_i\sim\ND(x_i^\top\beta,\sigma^2)$, where $\theta=(\beta^\top,\log \sigma^2)^\top$ is the stacked vector of model parameters. Write  $\tau=\log\sigma^2$.

First, consider the term $\mathbb{E}_{\phi(\theta,\mu,\Sigma)}[\log p(y\mid\theta,x)]$. We can decompose 
\begin{align*}
    \mu=\begin{pmatrix}
        \mu_\beta \\ \mu_\tau
    \end{pmatrix}\text{ and }  \Sigma=\begin{pmatrix}
        \Sigma_{\beta\beta} & \Sigma_{\beta\tau}\\ \Sigma_{\tau\beta} & \Sigma_{\tau\tau}.
    \end{pmatrix}
\end{align*}
We note that under $\theta\sim\ND(\mu,\Sigma)$ we can write $\beta\vert\tau$ as a normal distribution with mean $\mu_{\beta\vert\tau}(\tau)=\mu_\beta+\Sigma_{\beta\tau}\Sigma_{\tau\tau}^{-1}(\tau-\mu_\tau)$ and covariance $\Sigma_{\beta\vert\tau}=\Sigma_{\beta\beta}-\Sigma_{\beta\tau}\Sigma_{\tau\tau}^{-1}\Sigma_{\tau\beta}$. Then,
\begin{align*}
    &\mathbb{E}_{\phi(\theta,\mu,\Sigma)}[\log p(y\mid\theta,x)]=\sum_{i=1}^n-\frac{1}{2}\log(2\pi)-\frac{1}{2}\mathbb{E}_{\phi(\theta,\mu,\Sigma)}\left[\tau\right]-\frac{1}{2}\mathbb{E}_{\phi(\theta,\mu,\Sigma)}\left[\exp(-\tau)(y_i-x_i^\top\beta)^2\right]\\
    &=-\frac{n}{2}\log(2\pi)-\frac{n}{2}\mu_\tau-\frac{1}{2}\sum_{i=1}^n\mathbb{E}_{\phi(\theta,\mu,\Sigma)}\left[\exp(-\tau)(y_i-x_i^\top\beta)^2\right]\\
    &= -\frac{n}{2}\log(2\pi)-\frac{n}{2}\mu_\tau-\frac{1}{2}\exp(-\mu_\tau+\frac{1}{2}\Sigma_{\tau\tau})\sum_{i=1}^n\left\{(y_i-x_i^\top(\mu_\beta-\Sigma_{\beta\tau}))^2+x_i^\top\Sigma_{\beta\beta}x_i\right\}.
\end{align*}
Similarly for the score term,
 \begin{align*}
     &\int \phi(\Tilde{y};\Tilde{x}^\top\beta,\sigma^2)\phi(\theta;\mu,\Sigma)d\theta = \int\int  \phi(\Tilde{y};\Tilde{x}^\top\beta,\sigma^2) \phi(\beta;\mu_{\beta\vert\tau}(\tau),\Sigma_{\beta\vert\tau})d\beta\phi(\tau;\mu_{\tau},\Sigma_{\tau\tau})d\tau\\
     &= \int \phi(\Tilde{y};\Tilde{x}^\top \mu_{\beta\vert\tau}(\tau),\Tilde{x}^\top \Sigma_{\beta\vert\tau} \Tilde{x}+\exp(\tau))\phi(\tau;\mu_{\tau},\Sigma_{\tau\tau})d\tau\\
     &=\int \phi(\Tilde{y};\Tilde{x}^\top \mu_{\beta\vert\tau}(\mu_\tau+\sqrt{\Sigma_{\tau\tau}}w),\Tilde{x}^\top \Sigma_{\beta\vert\tau} \Tilde{x}+\exp(\mu_\tau+\sqrt{\Sigma_{\tau\tau}}w))\phi(w;0,1)dw\\
     &\approx \sum_{b=1}^B \gamma_b \phi\left(\Tilde{y};\Tilde{x}^\top \mu_{\beta\vert\tau}\left(\mu_\tau+\sqrt{\Sigma_{\tau\tau}}w_b\right),\Tilde{x}^\top \Sigma_{\beta\vert\tau} \Tilde{x}+\exp\left(\mu_\tau+\sqrt{\Sigma_{\tau\tau}}w_b\right)\right),
 \end{align*}
where $w_b$ and $\gamma_b$ are Gauss-Hermite quadrature points and weights respectively. We have $\mu_{\beta\vert\tau}\left(\mu_\tau+\sqrt{\Sigma_{\tau\tau}}w_b\right)=\mu_\beta+\Sigma_{\beta\tau}\sqrt{\Sigma_{\tau\tau}}^{-1}w_b$.

\paragraph{Poisson regression} 
Under the Poisson model $y_i\mid\theta\sim\PoisD(\exp(x_i^\top\theta))$
we have 
\begin{align*} \mathbb{E}_\phi(\theta;\mu,\Sigma)\left[\log p(y\mid\theta)\right]&=\sum_{i=1}^n\mathbb{E}_\phi(\theta;\mu,\Sigma)\left[y_ix_i^\top\theta-\exp(x_i^\top\theta)-\log(y_i!)\right]\\&=\sum_{i=1}^n y_ix_i^\top\mu-\exp(x_i^\top\mu+\frac{1}{2}x_i^\top\Sigma x_i)-\log(y_i!).
\end{align*}

\paragraph{Logistic regression}
Consider a logistic regression model 
\begin{equation*}
    \mathbb{P}(y_i=1)=\frac{1}{1+\exp(-x^\top\theta)}.
\end{equation*}
Then we approximate
\begin{align*}
    \mathbb{E}_{\phi(\theta,\mu,\Sigma)}[\log p(y\mid\theta)]&=\sum_{i=1}^n\mathbb{E}_{\phi(\theta,\mu,\Sigma)}\left[y_i\log \frac{1}{1+\exp(-x^\top\theta)} + (1-y_i)\log (1-\frac{1}{1+\exp(-x^\top\theta)})\right]
    \\&\approx \sum_{b=1}^B\gamma_b \bigg[y_i\log \frac{1}{1+\exp(-x^\top\mu-w_bx_i^\top\Sigma x_i)} \\&+ (1-y_i)\log (1-\frac{1}{1+\exp(-x^\top\mu-w_bx_i^\top\Sigma x_i)})\bigg],
\end{align*}
where $w_b$ and $\gamma_b$ are Gauss-Hermite quadrature points and weights respectively.

\section{Additional experiments} \label{app:applications}

\subsection{Simulations -- Linear regression}
We generate $n=1,000$ observations as 
\begin{align*}
    y_i &\sim \ND(x_i^3,\sigma^2),
\end{align*}
where the $x_i$ are uniform on $[-2,2]$. To this data we fit the misspecified linear regression model
\begin{align*}
    y_i&=\theta_1+x_i\theta_2+\varepsilon_i\\
    \varepsilon_i&\sim\ND(0,\sigma^2)\\
    \theta&\sim\ND(0,\tau^2I_2),
\end{align*}
where $\sigma^2=0.1$ and $\tau^2=100$ are known. 
We vary the tuning parameters $\beta$ for VGM-PVI and evaluate the predictive performance using the llpd on an additional hold-out test set with $n_\text{test}=10,000$. 

\begin{figure}[tbh]
    \centering
    \includegraphics[width=0.95\linewidth,keepaspectratio]{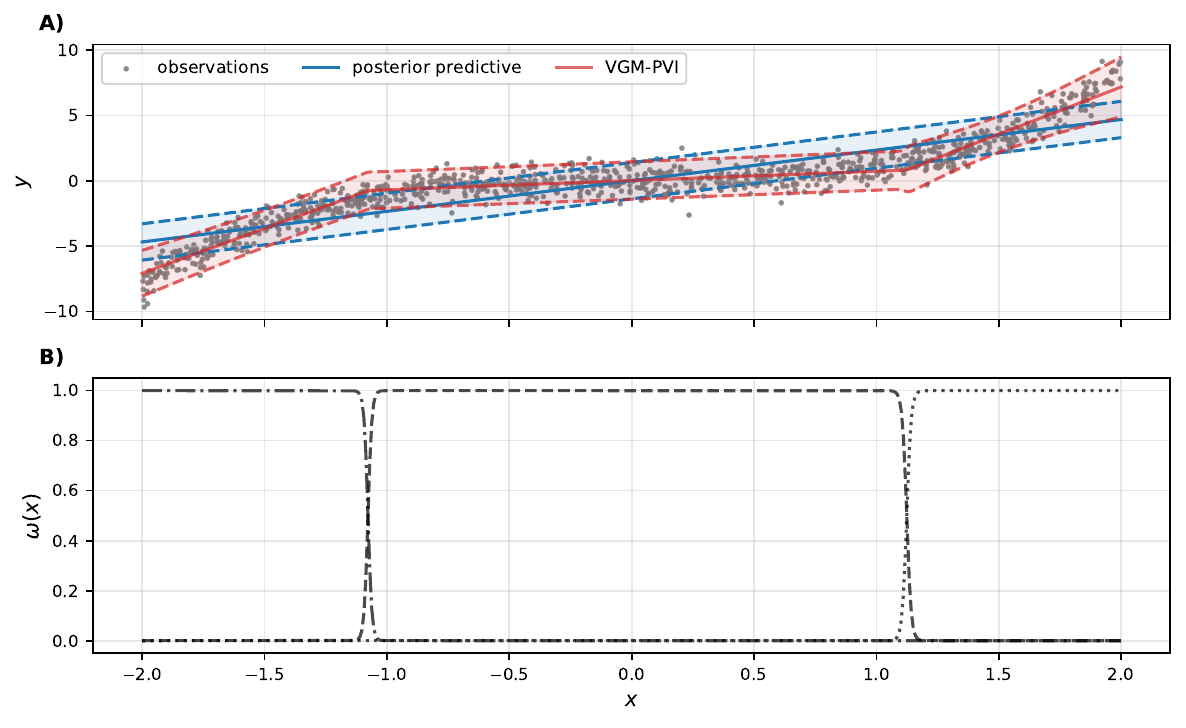}
    \caption{\small Simulation -- Linear regression. Panel \textbf{A)} shows the mean (bold) and a 95\% credible interval (dashed) derived by VGM-PVI under $\beta=0.01$ (red) and the true posterior (blue).  Panel \textbf{B)} shows the behaviour of the weights $\omega_k(x)$ under VGM-PVI. Different weights are differentiated by line-style.}
    \label{fig:sim_reg} 
\end{figure}

From previous discussion we know that for large values of $\beta$, VGM-PVI converges towards the true posterior predictive distribution, and that VGM-PVI matches the true DGP for small values of $\beta$. 
This is in-line with findings in \citet{lai+y25}. Figure~\ref{fig:sim_reg} A) compares the predictive distribution under VGM-PVI for the optimal value $\beta^*=0.01$ with the predictive distribution under the true posterior. The true posterior is incapable of capturing the non-linear relationship between $x$ and $y$ resulting in an unfavourable prediction model. In contrast to this, VGM-PVI is a covariate dependent mixture of linear models resulting in a piecewise linear approximation to the non-linear regression function. In particular, VGM-PVI identifies 3 regions in the covariate space and fits separate linear models to the data within each region. This can be seen by an investigation of the mixture weights given in Figure~\ref{fig:sim_reg} B). For all values of $\beta$ only a small number of mixture components is selected. In particular for large $\beta$ only a single component is selected, which is expected as the true posterior is a single Gaussian distribution. For the optimal value $\beta^\ast=0.01$, VGM-PVI selects three mixture components. Their activity corresponds to the kinks in the mean function of the DGP resulting in interpretable weights.

\subsection{Gaussian likelihood model -- IQ data}

This example illustrates PVI for a Gaussian likelihood model with potentially unknown variance. The dataset contains cognitive test scores of $n=434$ three and four-year-old children \citep{GelHil2007}. The response variable is the child's IQ $q_i$, and we have the mother's IQ, $m_i$, as well as a binary indicator indicating whether or not the mother completed high school, $h_i$, as covariables. After converting the IQ score to Z-scores, we consider the likelihood model
\begin{align*}
    q_i \sim \ND(\theta_0+\theta_1h_i+\theta_2m_i,\sigma^2),
\end{align*}
where we assume a prior $\theta=(\theta_0,\theta_1,\theta_2)^\top\sim \ND(0,I_3)$. In addition to a full Bayesian approach we also consider two plug-in estimators, $\widehat{\sigma}^2=\sigma_1^2,\sigma_2^2$, for $\sigma^2$ as this gives additional insight in the model extension provided by VGM-PVI. 
Expicitly, 
$\widehat{\sigma}^2=\sigma^2_1=(n-3)^{-1}\sum_{i=1}^n(q_i-\widehat{q}_{i,\text{LQE}})^2$, where $\widehat{q}_{i,\text{LQE}}$ is the predicted value under the least squares estimator for $\theta$, $i=1,\dots,n$. 
Also, $\widehat{\sigma}^2=\sigma_2^2=0.05\sigma_1^2$, so that the base model is bound to underestimate uncertainty. For the fully Bayesian approach, we assume the prior $\log \sigma^2\sim \ND(0,1)$ and $\log \sigma^2$ becomes an additional model parameter, so that $\theta=(\theta_0,\theta_1,\theta_2,\log\sigma^2)^\top$ in this case. We consider a randomly sampled subset of $80\%$ of the data for training and selection of $\beta$, while the remaining $20\%$ of the data are used for evaluation. Again, we select the optimal VGM-PVI by optimizing \eqref{eq:elpd} via BO. Table~\ref{tab:iq} reports both the in-sample and the out-sample llpd for all three scenarios. 

For $\widehat{\sigma}^2=\sigma_1^2$, VGM-PVI improves the conventional posterior only slightly, while we observe much greater improvement in the severely misspecified case $\widehat{\sigma}^2=\sigma_2^2$. In the third case, the conventional posterior performs better than VGM-PVI for both in- and out-sample predictions. 

\begin{table}[]
    \centering
    \begin{tabular}{l|cccccc}
    & \multicolumn{2}{c}{$\sigma^2=\sigma^2_1$} & \multicolumn{2}{c}{$\sigma^2=\sigma^2_2$}& \multicolumn{2}{c}{$\log\sigma^2\sim\ND(0,1)$}\\
         & OUT-llpd & IN-llpd&  OUT-llpd & IN-llpd& OUT-llpd & IN-llpd \\
         \hline\hline
      VGM-PVI   & -147.0766  & -542.9843 & -494.2149 & -1750.8228  & -154.2008 & -560.5294\\
      Posterior predictive   & -149.1583 & -547.825 & -810.4417& -3203.3109 & -149.2536 & -547.8748 \\
      \hline\hline
    \end{tabular}
    \caption{\small IQ data. llpd evaluated on the trainings data (IN-llpd), and the test data (OUT-llpd) for VGM-PVI and the conventional posterior predictive.}
    \label{tab:iq}
\end{table}

\subsection{Additional results: Gamma telescope data}
\begin{figure}[tbh]
    \centering
    \includegraphics[width=0.5\linewidth,keepaspectratio]{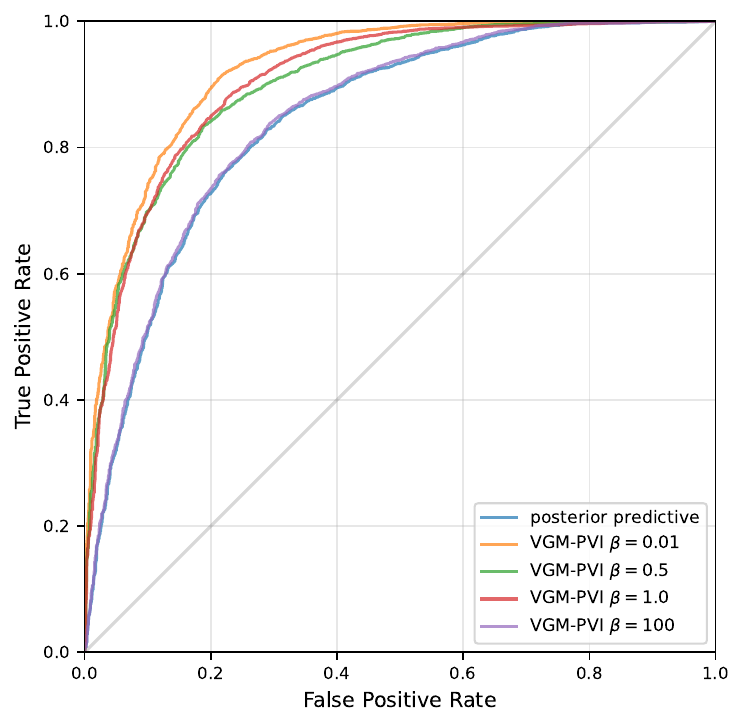}
    \caption{Gamma telescope. ROC curves on the hold-out test data for the conventional posterior and VGM-PVI with $\beta=0.01, 0.05, 1.0, 100$ (indicated by colour).}
    \label{fig:app:gamma}
\end{figure}

Figure~\ref{fig:app:gamma} shows the ROC curve for the Gamma telescope example. VGM-PVI improves on the base model for all values of $\beta$. For $\beta=100$, VGM-PVI performs similar to the posterior, while the best performance is achieved for $\beta=0.01$. 

